  \documentstyle[psfig]{kapproc} 

\upperandlowercase

\kluwerbib 

\begin{document}


\articletitle{Initial mass function and galactic chemical evolution models}


\chaptitlerunninghead{IMF and GCE models}





 \author{D. Romano\altaffilmark{1}, C. Chiappini\altaffilmark{2}, 
         F. Matteucci\altaffilmark{3}, M. Tosi\altaffilmark{1}}

\altaffiltext{1}{INAF--Osservatorio Astronomico di Bologna\\
                 Via Ranzani 1, I-40127 Bologna, Italy}
\altaffiltext{2}{INAF--Osservatorio Astronomico di Trieste\\
                 Via G. B. Tiepolo 11, I-34131 Trieste, Italy}
\altaffiltext{3}{Dipartimento di Astronomia, Universit\`a di Trieste\\
                 Via G. B. Tiepolo 11, I-34131 Trieste, Italy}

\email{donatella.romano@bo.astro.it,chiappini@ts.astro.it,matteucci@ts.astro.it,monica.tosi@bo.astro.it}

\begin{abstract}
In this contribution we focus on results from chemical evolution models for 
the solar neighbourhood obtained by varying the IMF. Results for galaxies of 
different morphological type are discussed as well. They argue against a 
universal IMF independent of star forming conditions.
\end{abstract}

\section*{Introduction}
Galactic chemical evolution (GCE) models are useful tools to understand how 
galaxies form and evolve. In particular, abundance and abundance ratio trends 
can be read as records of different evolutionary histories and interpreted in 
terms of the different time scales on which different objects evolve 
(\cite{mf89}; \cite{wst89}). Unfortunately, dealing with very complex and 
poorly known mechanisms such as mass accretion, star formation and stellar 
feedback, brings with it the need for many assumptions and parameters. As a 
consequence, GCE models are by no means unique (\cite{t88}). Therefore, it is 
worthwhile quantifying the uncertainties of GCE predictions arising from 
different treatments of the physical processes involved with structure 
formation and evolution. Here we concentrate on uncertainties due to different 
assumptions on the stellar IMF. By comparing the model predictions with the 
available data, we show that particular IMF slopes can be ruled out in the 
Galaxy, whilst a ``standard'' solar neighbourhood IMF is not suitable to 
describe the high metallicities observed in ellipticals.

\section{The IMF in the solar neighbourhood}

The most widely used functional form for the IMF is an extension of the 
``original mass function'' proposed by \cite{s55} to the whole stellar mass 
range, $\varphi(m) \propto m^{-1.35}$ for 0.1 $\le m/M_\odot \le$ 100. Besides 
this, multi-slope expressions (\cite{t80}; \cite{s86}; \cite{ktg93}; 
\cite{s98}) and a lognormal form for the low-mass part of the IMF ($m \le$ 1 
$M_\odot$; \cite{c03}) are considered here. In the latter case, for the 1--100 
$M_\odot$ stellar mass range we adopt a power-law form with an exponent $x$ = 
1.7.

\begin{figure}[b]
\centerline{\psfig{file=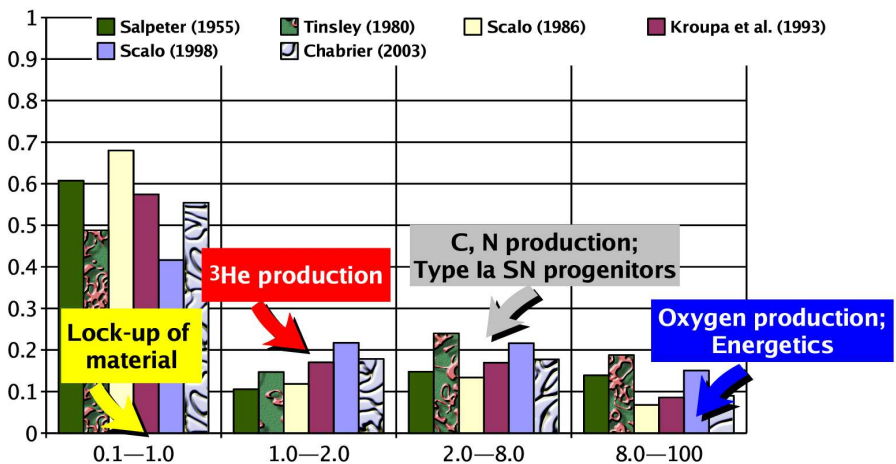,height=2.45in,width=4.7in}}
\caption{Fractional mass (y axis) falling in each given mass range (x axis) 
according to different IMF choices, for one single stellar generation.}
\end{figure}

\begin{figure}[ht]
\centerline{\psfig{file=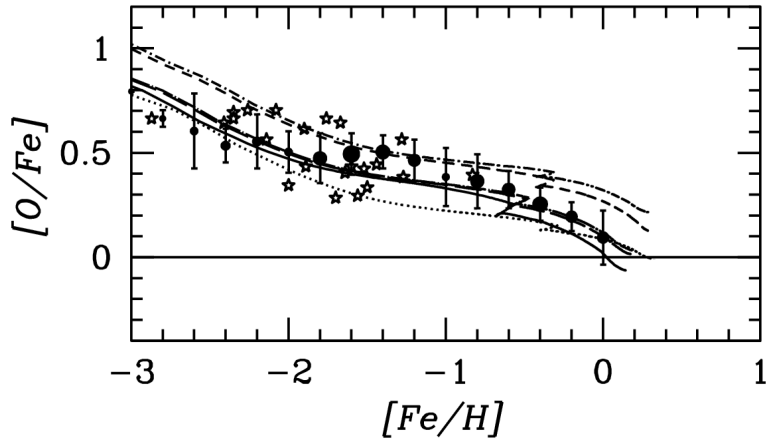,height=2.5in}}
\caption{[O/Fe] vs [Fe/H] in the solar neighbourhood as predicted by models 
adopting different IMFs (short-dashed line: Salpeter 1955; dotted line: 
\cite{t80}; solid line: \cite{s86}; long-dashed line: \cite{ktg93}; 
dot-short-dashed line: \cite{s98}; dot-long-dashed line: \cite{c03}). Data are 
mean [O/Fe] from [O\,{\footnotesize I}] lines in 0.2 dex metallicity bins 
(circles) and [O/Fe] values from infrared OH lines (stars; \cite{mb02}).}
\end{figure}

In Fig.~1 we show the fractional masses falling in specific mass ranges 
according to different IMF choices, for one single stellar generation. An 
example of what one gets when integrating over the Galactic lifetime, i.e. 
over many stellar generations, is given in Fig.~2, where we display the 
predicted behaviour of [O/Fe] vs [Fe/H] for all the IMFs listed above. We 
choose oxygen because of its well understood nucleosynthetic origin 
(\cite{f04}) and because the very high-quality data available for solar 
neighbourhood stars (\cite{mb02}) allow for a very meaningful comparison 
between model predictions and observations. Although theoretical errors of the 
order of 0.2--0.3 dex can be associated to model predictions owing to the 
uncertainties in the actual IMF form, it is apparent that both Salpeter's and 
Scalo's (1998) IMFs overproduce oxygen for most of the Galactic lifetime. 
Generally, models assuming Salpeter's, Tinsley's or Scalo's (1998) IMFs are 
found to predict far too high global metal abundances ($Z_\odot \sim$ 
0.024--0.033), especially if the most recent measurement of this quantity in 
the Sun is taken into account ($Z_\odot$ = 0.0126; \cite{a04}). The Scalo 
(1998) IMF also leads to overproduce $^3$He from the time of solar birth up to 
now, due to its high percentage of 1--2 $M_\odot$ stars (see Fig.~1). On the 
contrary, the remaining IMFs all guarantee a good agreement between the model 
predictions and the data (see Romano et al. 2004 for details).

Therefore, from simple GCE arguments we conclude that the IMF of field stars 
in the solar neighbourhood must contain less massive stars than the Salpeter 
one. An extrapolation of the Salpeter law to the high-mass domain is not 
suitable to explain the solar neighbourhood properties. This is not 
surprinsing; \emph{indeed, the Salpeter slope of $x$ = 1.35 was originally 
derived for stars less massive than 10 $M_\odot$}.

\section{The IMF in external galaxies}

The observational properties of dwarf galaxies are better explained by 
assuming a Salpeter-like stellar mass spectrum. This is true for both dwarf 
spheroidals (\cite{lm03}) and late-type dwarf galaxies (Romano, Tosi and 
Matteucci in preparation). On the other hand, the chemo-photometric properties 
of massive ellipticals at both low and high redshifts are better explained 
with an IMF slightly flatter than Salpeter's. In their pioneering work, 
\cite{ay87} showed that an IMF with a power index smaller than Salpeter, 
$\varphi(m) \propto$ $m^{-0.95}$ for 0.05 $\le m/M_\odot \le$ 60, gives an 
excellent fit to the observed colors of giant elliptical galaxies. We find 
that the chemo-photometric properties of local and high-redshift massive 
spheroids are well reproduced with an IMF slope even more similar to 
Salpeter's, i.e. $\varphi(m) \propto$ $m^{-1.25}$ for $m >$ 1 $M_\odot$ 
(\cite{rsmd02}).

In conclusions, our GCE models give us hints for (small) IMF variations with 
star forming conditions. Further studies have been presented at this workshop 
which seem to confirm our findings, from both a theoretical (e.g. C. Chiosi; 
P. Kroupa; L. Portinari, these proceedings) and an observational (e.g. S. 
Lucatello, these proceedings) point of view.

\begin{acknowledgments}
DR and MT wish to thank the organizers for the pleasant and interesting 
meeting.
\end{acknowledgments}

\end{document}